\documentclass[11pt, superscriptaddress, showpacs,showkeys]{revtex4}
\usepackage{amsmath,amsfonts,amssymb,graphicx,psfrag, epsfig,rotate}

\usepackage{amsmath,amsfonts,amssymb,graphicx,psfrag, epsfig,rotate}
\usepackage{eucal}
\usepackage{cmmib57}

\usepackage{colordvi}

\begin{document}

\title{ Thermodynamics of the $S=1$ spin ladder as a  
            composite $S=2$ chain model}

\date{\today}

\author{Onofre Rojas}
\affiliation{Departamento de Ci\^encias Exatas, Universidade Federal de
Lavras, Caixa Postal 37, CEP 37200-000, Lavras-MG,  Brazil.}

\author{E.V. Corr\^ea Silva}
\affiliation{Departamento de Matem\'atica e Computa\c c\~ao,
Faculdade de Tecnologia,
Universidade do Estado do Rio de Janeiro.
Estrada Resende-Riachuelo, s/n$^{\textit o}$,
Morada da Colina, CEP 27523-000,  Resende-RJ, Brazil.}

\author{S.M. de Souza}
\affiliation{Departamento de Ci\^encias Exatas, Universidade Federal de
Lavras, Caixa Postal 37, CEP 37200-000, Lavras-MG,  Brazil.}

\author{M.T. Thomaz}
\email[Corresponding author: ]{mtt@if.uff.br}
\affiliation{Instituto de F\'{\i}sica,
Universidade Federal Fluminense, Av. Gal. Milton Tavares de
Souza s/n$^{\textit o}$, CEP 24210-340, Niter\'oi - RJ, Brazil.}

\begin{abstract}

A special class of $S=1$ spin ladder hamiltonians, with second- neighbor
exchange interactions and  with
anisotropies in the $z$-direction, can be mapped onto one-dimensional
composite $S=2$ (tetrahedral $S=1$) models. We calculate the high
temperature expansion of the Helmoltz free energy for the latter class
of models, and show that their magnetization behaves closely to that of
standard XXZ models with a suitable effective spin $S_{eff}$, such
that $S_{eff}(1+S_{eff})=\langle \vec{\bf S}_i^2 \rangle$, where ${\bf
S}_i$ refers to the components of spin in the composite model. It is
also shown that the specific heat per site of the composite model, on
the other hand, can be very different from that of the effective spin
model, depending on the parameters of the hamiltonian.

\end{abstract}


\maketitle


Many materials have been described by low-dimensional spin
models\cite{lemmens} such as XXZ, ladder, tetrahedral, dimmer chain,
and mixed spin models. Different predictions are obtained from
one-dimensional and quasi-one-dimensional models that can be verified
experimentally; e.g., the $S=1/2$ antiferromagnetic Heisenberg chain
is a gapless model, whereas the even-legged antiferromagnetic
Heisenberg ladder model has a gap in its energy
spectrum\cite{dagotto}. 
Attention has also been drawn to composite spin models; for instance,
the $T=0$ phase diagram for the $S=(1/2 \oplus 1/2)$ model has been
studied by S\'olyom and Timonen\cite{solyom86,solyom89}. This
composite spin model is equivalent to the tetrahedral $S=1/2$ model
-- applied to the study of the properties of the tellurate materials $Cu_2
Te_2 O_5 Cl_5$ and $Cu_2 Te_2 O_5 Br_5$.
On the basis of  experimental results
it was argued that these materials could be appropriately
described by the noninteracting tetrahedral $S=1/2$
model\cite{totsuka}.  In Ref. \cite{PRB2003} 
we obtained the high temperature expansion
of the Helmholtz free energy of
the tetrahedral $S=1/2$ model up to order $\beta^5$.

A common feature shared by all composite spin models is that
the modulus of the spin at each site of the chain is not constant; 
instead, one has a random distribution throughout the chain.
The presence of spin $S=0$ in a given site of the chain 
can be interpreted as nonmagnetic impurity at that site\cite{PRB2003}.

By the appropriate design of molecules, it is possible to
obtain a variety of spin systems. 
One example is the organic compound BIP-TENO\cite{kato}, 
found to be a $S=1$ spin ladder. At $T=0$ this compound
exhibits a plateau-like anomaly at $1/4$ of the saturation 
magnetization, which has attracted much attention due to its full
quantum nature.
In Ref.\cite{sakai} a quasi one-dimensional spin model with second-
and third-neighbor antiferromagnetic exchange interactions 
suitably explains the mechanism of the $1/4$ magnetization plateau of this  
material; it does not include any anisotropy in the $z$-direction, though.
Numerical analysis has been applied to study
its phase diagram\cite{okamoto} at $T=0$; even in the absence of 
third-neighbor interactions the magnetization plateau is shown to be present.

In Ref. \cite{chain_m} we presented a method to 
calculate the coefficients of
the cummulant expansion of the Helmholtz free energy, in the
thermodynamic limit, of any chain model whose hamiltonian 
satisfy periodic boundary conditions, possess invariance under spatial 
discrete translations and include interactions between nearest neighbors. 

\vspace{0.3cm}

By a suitable choice of constants in the hamiltonian of
Ref.\cite{sakai}, the $S=1$ ladder with second-neighbor interactions
($S=1$ tetrahedral model, cf. model $B$ in Ref.\cite{niggemann97}; at
each site, spin- 1/2 degrees of freedom are replaced by those of a 
fundamental spin- 1) can be mapped  onto a family of one-dimensional
 composite $S=2$ chain models.
In the present work, we increase the parameter space of the model in
Ref.\cite{sakai} by introducing anisotropies in the $z$-direction, 
calculate the high-temperature expansion of its Helmholtz free energy 
and compare its thermodynamics, in the same regime of temperature, 
to the  standard (irreducible representation ) spin- $S$ XXZ model.
The high-temperature expansion (or $\beta$-expansion, where
$\beta=1/kT$, $k$ is the Boltzmann constant and $T$ is the absolute
temperature) of the Helmholtz free energy for the composite $S=2$
XXZ model is obtained by applying the method presented in
Ref.\cite{chain_m}. From the analytical results we may obtain the
thermodynamic functions of the  model at high temperatures; in particular, 
the magnetization ${\cal M}$ as a function of the external magnetic 
field $h$ and the specific heat per site $C_L$ as a function of $\beta$.

\vspace{0.7cm}

The hamiltonian  of the quasi-one-dimensional spin model is

\begin{eqnarray}
 H_{Q-1D} &=& \sum_{i=1}^{N} \big\{ J_0 [ (\sigma_i, \tau_i)_{\Delta_0} 
 + \frac{1}{2} (\Delta_0 -1) ( (\sigma_i^z)^2 \otimes {\bf 1}_\tau  
 + {\bf 1}_{\sigma} \otimes (\tau_i^z)^2) ] 
                  \nonumber \\
&+& 
 J[ (\sigma_i, \sigma_{i+1})_\Delta \otimes {\bf 1}_\tau +
  (\sigma_i, \tau_{i+1})_\Delta 
+  (\tau_i, \sigma_{i+1})_\Delta
+ {\bf 1}_{\sigma} \otimes  (\tau_i, \tau_{i+1})_\Delta]
      \nonumber \\
%
%
& - & h (\sigma_i^z\otimes {\bf 1}_\tau + 
	      {\bf 1}_\sigma\otimes \tau_i^z) \big\},
	         \label{1}
 \end{eqnarray}

\noindent  and it is subject to periodic boundary conditions. 
We use the same  notation as in  Ref.\cite{PRB2003}:
$(A_l, B_k)_\Delta \equiv A_l^x\otimes B_k^x + A_l^y\otimes B_k^y
+ \Delta A_l^z\otimes B_k^z$, with $A_l \equiv (A_l^x, A_l^y, A_l^z)$
and $B_k \equiv (B_k^x, B_k^y, B_k^z)$, introducing the anisotropy
in the $z$-direction. For $\Delta_0 =1$ and $\Delta=1$,  (\ref{1})
equals to the sum of hamiltonians (2) and (3) of reference \cite{sakai}  
for the $S=1$ ladder with second-neighbor
exchanges ($J_3 = 0$),  for the special case $J_1 = J_2$. 
The distinct $S=1$ variables $\sigma_i$ and $\tau_i$ are 
related to the $\rho$- and $r$-lines of the dumb-bell, respectively 
(cf. Fig.1 of Ref.\cite{PRB2003}).

We define the composite  spin operators $\vec{S}_i$  at the $i$-th site as 
$ \vec{S}_i = \vec{\sigma}_i \otimes {\bf 1}_\tau + {\bf 1}_\sigma 
\otimes \vec{\tau}_i$.
Here, ${\bf 1}_\sigma$ and ${\bf 1}_\tau$ are the identity
operators in $\sigma$- and $\tau$-space, respectively.
Using the composite spin  $\vec{\bf S}_i$,  the tetrahedral 
$S=1$ hamiltonian (\ref{1}) is rewritten as a 
composite $S=2$ chain hamiltonian
\begin{eqnarray} 
{\bf H}_{Q-1D} & =&    \sum_{i=1}^N \left[ -2J_0 {\bf 1}
+ g ({\bf S}_i, {\bf S}_i)_1 + 
J \left( {\bf S}_{i}^{+} {\bf S}_{i+1}^{-} +  
{\bf S}_{i}^{-} {\bf S}_{i+1}^{+}  + 
 \Delta  {\bf S}_{i}^{z} {\bf S}_{i+1}^{z} \right)  \right.  \nonumber\\
 && \hspace{1cm} \left.
- h {\bf S}^{z}_{i} + d ({\bf S}^{z}_{i})^2 \right] ,  
      \label{2}
\end{eqnarray}

\noindent where $g \equiv \frac{J_0}{2}$, $d \equiv
\frac{J_0}{2}(\Delta_0 -1)$, $S^{\pm}_i \equiv \frac{1}{\sqrt{2}}
(S_i^x \pm i S_i^y)$, and ${\bf 1} $ is the identity operator,
represented by a $9\times 9$ identity matrix. The 
block matrix representations of the composite spin operators in 
(\ref{2}), in the basis of eigenstates of ${\bf S}_{i}^{z}$ 
 and $\vec{\bf S}_{i}^{2}$, are

	 \begin{eqnarray}
	S_i^{z} &=&  \left[\begin{array}{ccc}
	\Sigma^z_{(2)} & 0 & 0 \\
	0 & \Sigma^z_{(1)} & 0 \\
	0 & 0 & \Sigma^z_{(0)}
	\end{array}
	 \right]
	\hspace{0.5cm} {\rm and} \hspace{0.5cm}
	S_i^+ =  \left[\begin{array}{ccc}
	\Sigma^+_{(2)} & 0 & 0 \\
	0 & \Sigma^+_{(1)} & 0 \\
	0 & 0 & \Sigma^+_{(0)}
	\end{array}
	 \right],  \nonumber \\
	      \label{3}
	\end{eqnarray}
	
\noindent where the $\Sigma$ square matrices refer to the different
	spin sectors,

 \begin{eqnarray}
	\Sigma^z_{(2)} =  \left[\begin{array}{cccccc}
	-2 & 0 & 0 & 0 & 0  \\
	0 & -1 & 0 & 0 & 0  \\
	0 & 0 & 0 & 0 & 0  \\
	0 & 0 & 0 & 1 & 0  \\
	0 & 0 & 0 & 0 & 2 
	\end{array}
	 \right] &\ ,\ &
	\Sigma^+_{(2)} =  \left[\begin{array}{ccccc}
	0 & 0 & 0 & 0 & 0  \\
	\sqrt{2} & 0 & 0 & 0 & 0  \\
	0 & \sqrt{3} & 0 & 0 & 0  \\
	0 & 0 & \sqrt{3} & 0 & 0  \\
	0 & 0 & 0 & \sqrt{2} & 0 
	\end{array}
	 \right],  \nonumber \\
	\Sigma^z_{(1)} =  \left[\begin{array}{ccc}
	-1 & 0 & 0  \\
	0 & 0 & 0   \\
	0 & 0 & 1 
	\end{array}
	 \right] &\ ,\ &
	\Sigma^+_{(1)} =  \left[\begin{array}{ccc}
	0 & 0 & 0 \\
	1 & 0 & 0 \\
	0 & 1 & 0 
	\end{array}
	 \right], \nonumber \\
	\Sigma^z_{(0)} =  \left[\begin{array}{c}
	0 
	\end{array}
	 \right] &\ ,\ &
	\Sigma^+_{(0)} =  \left[\begin{array}{c}
	0 
	\end{array}
	 \right] .
	\end{eqnarray}

\noindent We point out that the operator $\vec{\bf S}_{i}^{2}$ is a
constant of the motion, although the hamiltonian (\ref{2}) can be
interpreted as mixture of 3 kinds of spin ($S= 0, 1$ and $2$),
randomly distributed along the chain, with its probability depending
on the constants of the hamiltonian and on the temperature.

An interesting  limit  of the hamiltonian (\ref{2}) 
is $g= J_0/2 \rightarrow 0$, for finite values of $d \equiv \frac{J_0 \Delta_0}{2}$.
In this limit  it acquires a single-ion anisotropy term.

As in  Ref.\cite{PRB2003},  the presence 
of a spin  $S=0$   in a site of the chain can be interpreted as nonmagnetic
impurity at that site. The model thus encompasses the presence of randomly 
distributed impurities along the chain, without hindering the application of the method 
of  Ref. \cite{chain_m}, since  the hamiltonian  (\ref{2}) has only 
nearest-neighbor interactions and is invariant under 
spatial  translations. 

Appendix \ref{A} presents the $\beta$-expansion of  the Helmholtz free 
energy $\mathit{W}_\mathit{Q-1D}$  of the composite 
$S=2$ XXZ model,  up to order $\beta^6$. Although this expansion
 has a large number of  terms, it is easily  differentiaded  by any 
symbolic computer  language,   yielding the thermodynamic functions  of the tetrahedral 
$S=1$   model.

For finite temperatures the results are insensitive to the sign of $J$. 
If we redefine the parameters of the hamiltonian (\ref{2})
in units of $J$,  the expression (\ref{A1}) of $\mathit{W}_\mathit{Q-1D}$
becomes an expansion in $(\beta J)$.  For the sake of simplicity, we 
choose $J=1$ in what follows.

As   we are discussing the case  of   the  quasi one-dimensional
model (the tetrahedral  $S=1$   model)  onto  a chain model, it is interesting  to compare
the high-temperature thermodynamic properties of the latter,  to those of the
standard spin-$S$ XXZ model, for distinct values of $S$.

The free energy $\mathit{W}_\mathit{Q-1D}$ and 
the mean value of the squared norm of spin $\langle\vec{\bf S}_i^2\rangle$,  for
the $S=1$ ladder model with second-neighbor exchange interactions, are related by
$\langle \vec{\bf S}_i^2\rangle = \frac{\partial \mathit{W}_\mathit{Q-1D}}{\partial g}$, 
for $g \equiv \frac{J_0}{2}$ and  $d \equiv \frac{J_0}{2}(\Delta_0 -1)$.
{$\langle \vec{\bf S}_i^2\rangle$ is a function of 
temperature and of the parameters of the hamiltonian (\ref{2}).}
At $\beta=0$,  the universal value   $\langle \vec{\bf S}_i^2\rangle =4$ 
is obtained, since  in this limit of temperature we have $N$ independent
 spins with equal   probability to be at $S=0,1$ or $2$. 

\begin{figure}
\includegraphics[width=11cm,height=16cm,angle=-90]{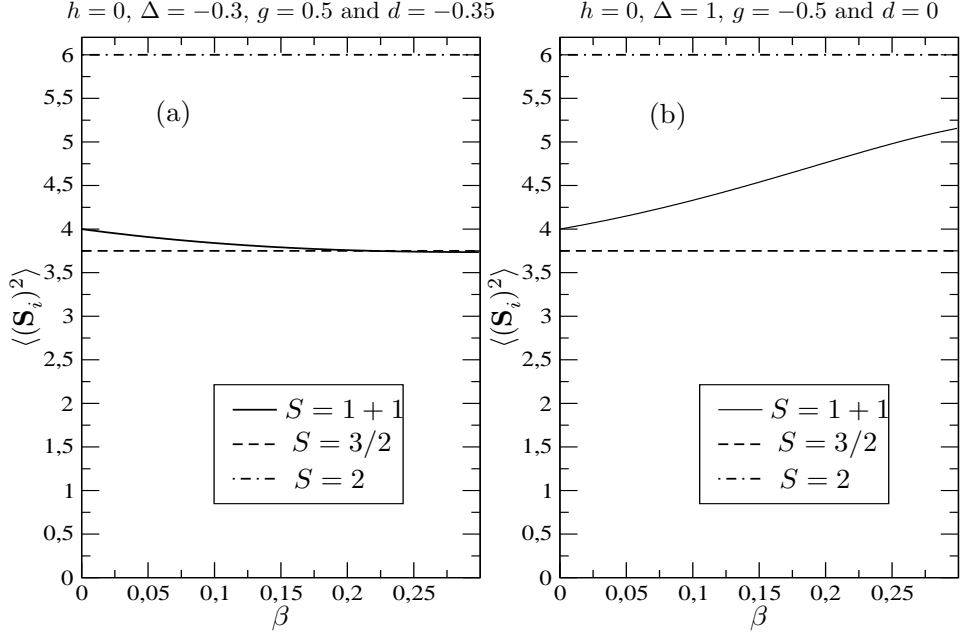}
\caption{Expectation values of the squared norm of spin 
$\langle\vec{\bf S}_i^2\rangle$ as a function of $\beta=1/kT$ for the composite 
$S=2$ model (solid lines) and the standard spin-$S$ model, with $S= 3/2$ (dashed lines)
and $S=2$ (dot-dashed lines), for a vanishing magnetic field $h$.
The values of parameters of the hamiltonian (\ref{2}) have been
chosen as (a) $\Delta= -0.3$,  $g=0.5$ and $d= -0.35$ and
(b) $\Delta= 1, g= -0.5$ and $d= 0$ (no	anisotropy in the $z$-direction).}
\label{fig1}
\end{figure}

Fig.\ref{fig1} shows  $\langle \vec{\bf S}_i^2\rangle$ as a function of 
$\beta$ for two sets of values for the parameters of the composite model.
$\langle \vec{\bf S}_i^2\rangle$ is not very sensitive to the value of $h$, 
within the range of $h$ where its $\beta$-expansion is a good approximation.
We let  $h=0$ in  Fig.\ref{fig1};   in Fig.\ref{fig1}a we have 
$ \Delta = -0.3, g= 0.5$ and  $d= -0.35$; and in 
Fig.\ref{fig1}b  $ \Delta = 1,  g=  -0.5$ and  $d= 0$. 
For the sake of comparison, 
Fig.\ref{fig1} also shows straight horizontal lines
that correspond to the squared norms of spin per site 
for the  standard   XXZ model with $S= 3/2$ and $2$.




\begin{figure}
\includegraphics[width=11cm,height=16cm,angle=-90]{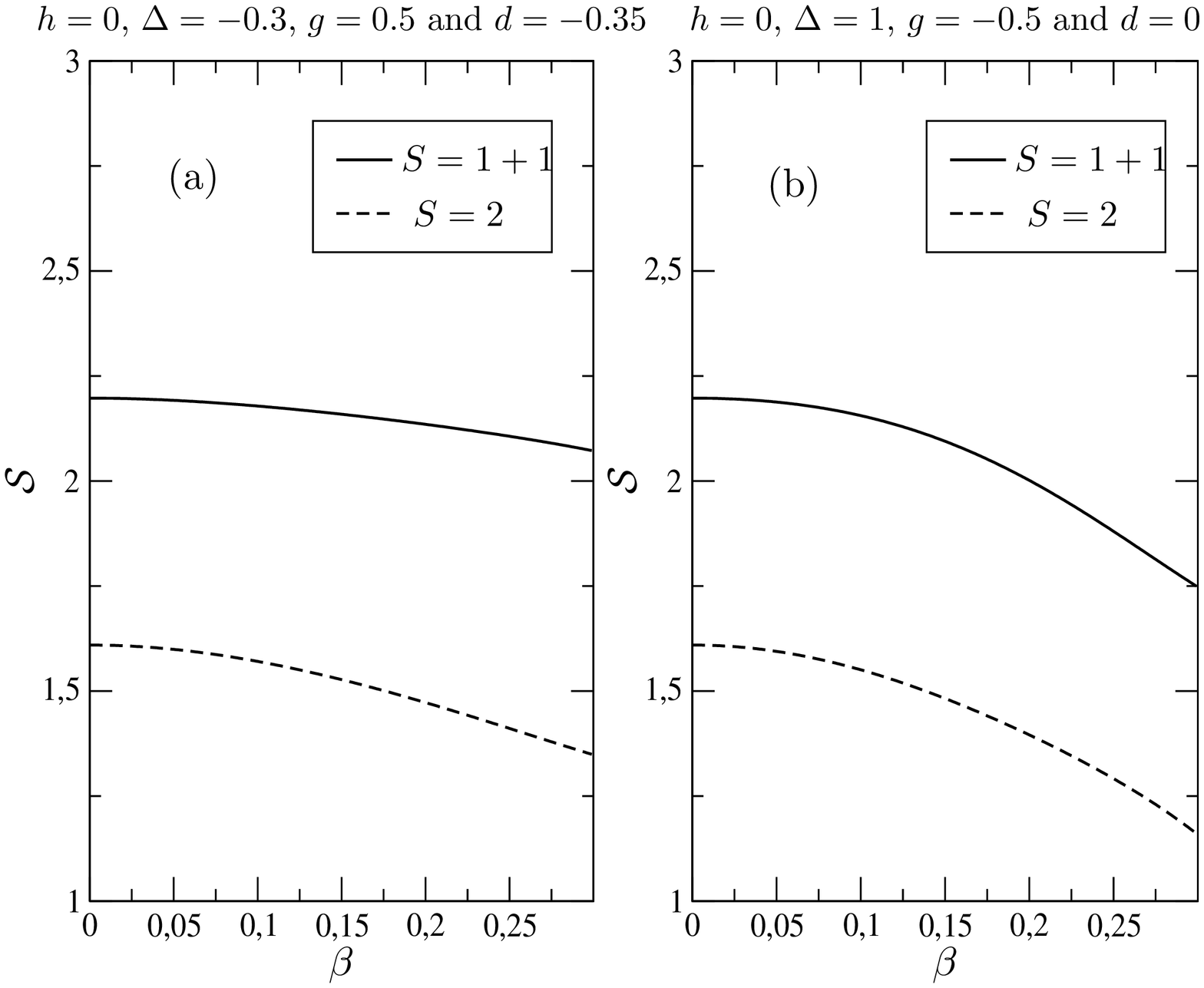}
\caption{Comparison of the entropy ${\cal S}$ as a function of $\beta=1/kT$ for 
the composite $S=2$ XXZ model (solid line) and the irreducible $S=2$ XXZ model (dashed line). 
The values of parameters in Figs.\ref{fig2}a and \ref{fig2}b 
are the same as in Figs.\ref{fig1}a and \ref{fig1}b, respectively.}
\label{fig2}
\end{figure}

Fig.\ref{fig2}  shows the entropy  ${\cal S}$   as a function of  $\beta$,
in units  of   Boltzmann's constant 
{\mbox{(${\cal S} = \beta^2 \frac{\partial {\cal W}}{\partial \beta}$)}}, for the 
same set of  parameter values  as in  Fig.\ref{fig1}. In both cases the entropy
of the tetrahedral  $S=1$   model is higher than the entropy of the 
composite $S=2$ XXZ model.
This result agrees with the fact that the number of degrees of
freedom of the composite $S=2$ model is {\it higher} than
that of the standard (irreducible) $S=2$ XXZ model.




\vspace{0.3cm}

In reference \cite{spin-S} Rojas {\it et al}. derived the high temperature
expansion of the Helmholtz free  energy ${\cal W}_\mathit{S} $ of the standard 
spin-$S$  XXZ  model, for arbitrary value of 
$S$ ($S$ being semi-integer or integer), up to order $\beta^6$. 
From their result, we obtain the expansion
of the magnetization (${\cal M}$) of this family of models 
(${\cal M} = -\frac{\partial {\cal W}}{\partial h}$). 
Appendix \ref{B} shows this expansion up to order $\beta^6$. 
Postulating the validity of expansion (\ref{B1}) for {\it real} positive
 values of $S$, we  define an {\it effective}   spin value 
$S_{eff}$   such that 
$S_{eff}(S_{eff}+1)  = \langle\vec{\bf S}_i^2\rangle$
where $\langle\vec{\bf S}_i^2\rangle$ relates to the composite $S=2$ model.

\begin{figure}
\includegraphics[width=11cm,height=16cm,angle=-90]{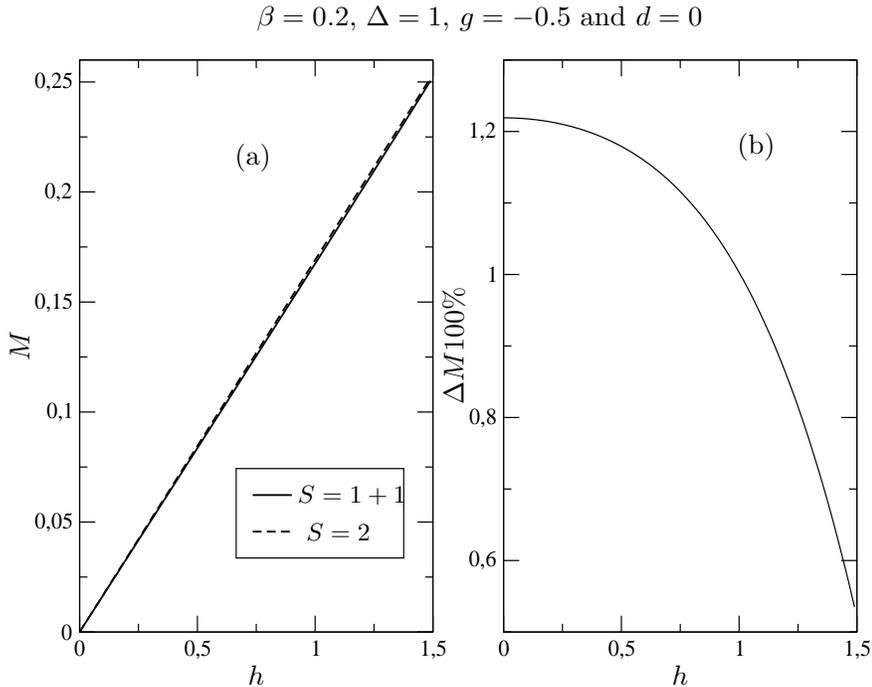}
\caption{The solid line in Fig.\ref{fig3}a shows the magnetization ${\cal M}$ as a function of
the external magnetic field $h$ at $\beta = 0.2$ and for the following values of
parameters of the composite 
$S=2$ XXZ model(\ref{2}): $\Delta =1$, $g= -0.5$
and $ d=0$.  The dashed line  is the
magnetization curve for the effective $S_{eff}$ XXZ model. 
Fig.\ref{fig3}b shows the percent error of the two curves in Fig.\ref{fig3}a 
as a function of $h$.}
\label{fig3}
\end{figure}

Obviously, the effective spin depends on the   values
of the parameters of the theory   (see  Fig.\ref{fig1}) 
and, in general, it is neither integer nor semi-integer.
Fig.\ref{fig3}a compares  the magnetization of the tetrahedral 
$S=1$   model and  that of the effective XXZ model 
at $\beta= 0.2$,   for $\Delta = 1, g= -0.5$ and $d=0$.  
Fig.\ref{fig3}b shows  the  percent error  of the two curves 
in Fig.\ref{fig3}a.  Such error is less than 1.3\% for $h$ $\in$ $[0, 1.4]$.




\begin{figure}
\includegraphics[width=11cm,height=16cm,angle=-90]{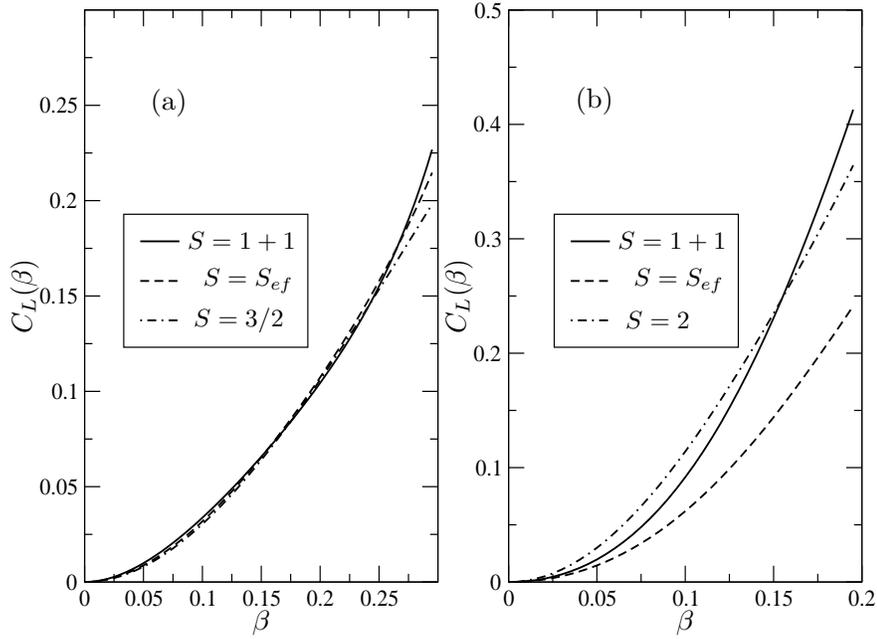}
\caption{(a) The specific heat per site $C_L$ as a function of
$\beta$ for the composite $S=2$ model (solid line), the $S=3/2$ XXZ
model (dot-dashed line), and the effective $S_{eff}$ XXZ model (dashed
line), for $h=0$, $d= -0.35$, $\Delta = -0.3$ and $g= 0.5$. (b) The
same, for the composite $S=2$ model (solid line), the standard $S=2$
XXZ model (dot-dashed line) and the corresponding effective
irreducible spin $S_{eff}$ model (dashed line) for $h=0$, $d=0$,
$\Delta = 1$ and $g= -0.5$. (The values of $S_{eff}$ are different, in
each case.)}
\label{fig4}
\end{figure}

This similarity of behavior is not shared, in general, by the specific heat per site 
$C_L$ ($C_L = -\beta^2 \frac{\partial^2  (\beta {\cal W})} {\partial \beta^2}$), as 
shown in Fig.\ref{fig4}.

Fukushima {\it et al}.\cite{fukushima}  calculated the $\beta$-expansion of the specific 
heat per site of the spin-$S$ XXZ model, for arbitrary (semi-integer or integer) values of $S$, 
up to order  $\beta^{11}$.  Proceeding similarly as done for the  magnetization, we 
postulate   the validity of  their expansion  of the specific heat per 
site   to any real positive value of   $S$.
Fig.\ref{fig4}a shows the specific heat per site $C_L$  
as a function of $\beta$ for the tetrahedral $S=1$, 
the standard $S=3/2$  and the effective $S_{eff}$ XXZ models; 
parameter values $ h=0$, $\Delta = -0.3$, 
$g= 0.5$ and $d = -0.35$ are the same as in Fig.\ref{fig1}a.
At high temperatures the curves are close:
for $\beta \in [0.13, 0.27]$ the percent error is smaller than 5\%, but
for $\beta\sim 0$ it goes up to 27\%.

Fig.\ref{fig4}b was obtained for the same values of parameters used in
Fig.\ref{fig1}b, that is, $h=0$, $\Delta = 1$, $g= -0.5$ and $d=0$.
The $C_L$ curves for the effective and the composite
$S=2$ models are very different: the percent error for 
$\beta \in [0, 0.2]$ varies from 20\% at $\beta =0$, to 41.8\% at $\beta = 0.2$.
Although there are $S=0$ and $S=1$ sites in a composite  
$S=2$ XXZ chain, its specific heat is larger than that of a standard $S=2$ chain,
for $\beta > 0.15$.




{\vspace{0.6cm}}

In summary, we applied the method of Ref.\cite{chain_m} to calculate
the high temperature expansion ($\beta$-expansion), up to order
$\beta^6$, of the Helmholtz free energy of a special class of quasi
one-dimensional models, the $S=1$ ladders with second neighbor 
exchange interactions (the tetrahedral $S=1$
models\cite{niggemann97}). We have increased the parameter space of
this class of hamiltonians by introducing anisotropies along the
$z$-direction. This class of quasi one-dimensional models can be mapped
onto one-dimensional composite $S=2$ XXZ models, with
spins $S=0,1$ and $2$ randomly distributed along the chain.
 At high temperatures, the thermodynamic properties of the 
 {\it  composite} $S=2$ XXZ model and the {\it standard} spin-$S$ XXZ model
are expected to differ, in principle; this is the case of the specific
heat. However, a somewhat surprising result is that the magnetization
of the tetrahedral $S=1$ model follows that of an effective
irreducible spin $S_{eff}$ model very closely (up to a percent error
smaller than 2\%). The effective spin is such that $S_{ef}(S_{ef}+1) =
\langle \vec{\bf S}_i^2 \rangle$, and $\langle \vec{\bf S}_i^2
\rangle$ is the expectation value of the squared norm of spin per site
of the composite $S=2$ XXZ model. The function $\langle \vec{\bf
S}_i^2 \rangle$ is a real-valued continuous function; $S_{eff}$ may
not be either integer or semi-integer.

\begin{acknowledgments}

O. R. thanks FAPEMIG for financial support.
S.M. de S. thanks FAPEMIG and CNPq for partial financial support.
M.T.T. thanks CNPq and FAPERJ for partial financial
support.

\end{acknowledgments}

\newpage

\appendix 

\section{The Helmholtz free energy of the  tetrahedral $S=1$ model }  \label{A}

The following expression for the high-temperature expansion up to
order $\beta^6$ of the Helmholtz free energy refers to the composite
$S=2$ model (that is, the tetrahedral $S=1$ model), whose dynamics is
driven by the hamiltonian (\ref{2}). It is valid for any real values
of $J_0, g, J, \Delta, d$ and $h$. (Observe that these values may be
positive, null or negative). When $g = \frac{J_0}{2}$ and $d =
\frac{J_0}{2}(\Delta_0 -1)$, it yields the free energy of the
hamiltonian (\ref{1}), for the $S=1$ ladder model with second-neighbor
interactions.

\begin{equation}\label{A1}
{\cal W}_{Q-1D} = 
	- \frac{\mathrm{ln}(9)}{\beta}
	+ W_0
	+ W_1 \ \beta
	+ W_2 \ \beta^2
	+ W_3 \ \beta^3
	+ W_4 \ \beta^4
	+ W_5 \ \beta^5
	+ W_6 \ \beta^6
	+ {\cal O} (\beta^7),
\end{equation}
where
{\footnotesize
\begin{equation} 
W_0 = - 2\,\mathit{J_0} + { \frac {4\,d}{3}}  + 4\,g 
\end{equation}

\begin{equation} 
W_1 =  - { \frac {16\,d\,g}{9}}  
	- { \frac {8\,g^{2}}{3}}  
	- { \frac {8\,J^{2}\,\Delta ^{2}}{9}}  
	- { \frac {2\,h^{2}}{3}}  - { \frac {10\,d^{2}}{9}}  
	- { \frac {16\,J^{2}}{9}}
\end{equation}

\begin{eqnarray} 
W_2 &=& { \frac {32\,d^{2}\,g}{27}}
 - { \frac {8\,d\,g^{2}}{9}}    
  + { \frac {64\,J^{2}\,\Delta ^{2}\,g}{27}} 
 - { \frac {8\,g^{3}}{9}}  + { \frac {8\,h^{2}\,g}{9}} 
 + { \frac {10\,d\,h^{2}}{9}}  + { \frac {46\,d^{3}}{81}}  \nonumber \\
 && \nonumber \\
 &&  - { \frac {4\,J^{3}\,\Delta }{9}}  
 + { \frac {16\,J\,\Delta \,h^{2}}{9}}  
 + { \frac {128\,g\,J^{2}}{27}}   
  + { \frac {80\,d\,J^{2}\,\Delta ^{2}}{27}} 
 - { \frac {16\,d\,J^{2}}{27}} 
\end{eqnarray}

\begin{eqnarray} 
W_3 &=& 
 - { \frac {146\,J^{4}}{81}}  
 + { \frac {512\,g\,d\,J^{2}}{81}}  
 + { \frac {32\,J^{3}\,\Delta \,d}{81}}  
 - { \frac {32\,J^{2}\,\Delta ^{2}\,g^{2}}{81}}  
  + { \frac {116\,J^{4}\,\Delta ^{2}}{81}}  \nonumber \\
 &&
- { \frac {128\,h^{2}\,J\,\Delta \,g}{27}}  
- { \frac {104\,J^{2}\,\Delta ^{2}\,h^{2}}{27}}  
+ { \frac {32\,g\,J^{3}\,\Delta }{27}}  
- { \frac {32\,h^{2}\,d\,g}{27}}  
- { \frac {23\,h^{2}\,d^{2}}{27}}    \nonumber \\
&&
 + { \frac {64\,d\,g^{3}}{27}}  
- { \frac {64\,J^{2}\,\Delta ^{2}\,d\,g}{9}}  
+ { \frac {4\,h^{2}\,g^{2}}{9}}  
+ { \frac {20\,h^{2}\,J^{2}}{27}}  
- { \frac {34\,J^{4}\,\Delta ^{4}}{27}}  
- { \frac {128\,J^{2}\,\Delta ^{2}\,d^{2}}{27}}   \nonumber \\
&&
 - { \frac {64\,g^{2}\,J^{2}}{81}}  
+ { \frac {140\,d^{2}\,J^{2}}{81}}  
+ { \frac {140\,g^{2}\,d^{2}}{81}}  + { \frac {16\,g^{4}}{9}}  
+ { \frac {32\,g\,d^{3}}{81}}  + { \frac {h^{4}}{18}} 
- { \frac {160\,h^{2}\,J\,\Delta \,d}{27}}    \nonumber \\
&&  + { \frac {17\,d^{4}}{162}} 
\end{eqnarray}

\begin{eqnarray} 
W_4 &=& 
{ \frac {664\,J^{3}\,\Delta ^{3}\,h^{2}}{81}}  
+ { \frac {448\,J^{2}\,\Delta ^{2}\,d\,h^{2}}{27}}  
+ { \frac {256\,J\,\Delta \,h^{2}\,d^{2}}{27}}  
- { \frac {17\,d^{3}\,h^{2}}{81}}    \nonumber \\
&&
 - { \frac {178\,d\,J^{4}}{243}}  
- { \frac {58\,J^{5}\,\Delta }{243}}  
- { \frac {19\,h^{4}\,d}{54}}  
- { \frac {29\,J^{5}\,\Delta ^{3}}{243}}  
- { \frac {44\,J^{2}\,d^{3}}{81}}  
- { \frac {100\,d^{2}\,J^{3}\,\Delta }{243}}     \nonumber \\
&&
 - { \frac {1538\,d\,J^{4}\,\Delta ^{2}}{243}}  
- { \frac {16\,J\,\Delta \,h^{4}}{27}}  
- { \frac {172\,d\,J^{2}\,h^{2}}{81}}  
+ { \frac {784\,J^{2}\,\Delta ^{2}\,d^{3}}{243}}  
+ { \frac {428\,J^{4}\,\Delta ^{4}\,d}{81}}    \nonumber \\
&&
 - { \frac {280\,J^{3}\,\Delta \,h^{2}}{81}} 
 - { \frac {191\,d^{5}}{486}} 
 - { \frac {32\,g\,d\,J^{3}\,\Delta }{243}}  
 - { \frac {64\,g^{2}\,J^{2}\,\Delta ^{2}\,d}{27}}  
 + { \frac {352\,J^{2}\,\Delta ^{2}\,h^{2}\,g}{27}}    \nonumber \\
&& 
 + { \frac {544\,g\,J^{2}\,\Delta ^{2}\,d^{2}}{81}}  
+ { \frac {64\,J\,\Delta \,h^{2}\,g^{2}}{81}}  
+ { \frac {320\,J^{4}\,\Delta ^{4}\,g}{81}}  
- { \frac {16\,J^{3}\,\Delta \,g^{2}}{81}}  
- { \frac {140\,d\,h^{2}\,g^{2}}{81}}     \nonumber \\
&&
 - { \frac {928\,J^{4}\,\Delta ^{2}\,g}{81}} 
 - { \frac {128\,J^{2}\,\Delta ^{2}\,g^{3}}{27}}  
 + { \frac {200\,g^{4}\,d}{81}}  
 - { \frac {352\,g\,d^{4}}{243}}  - { \frac {8\,h^{4}\,g}{27}}  
 - { \frac {428\,g^{2}\,d^{3}}{243}}   \nonumber \\
&& 
 - { \frac {16\,d^{2}\,g^{3}}{27}} 
- { \frac {32\,h^{2}\,g^{3}}{27}}  
+ { \frac {176\,g\,J^{4}}{81}}  - { \frac {256\,J^{2}\,g^{3}}{27}}  
+ { \frac {40\,g^{5}}{27}}  
- { \frac {16\,g\,d^{2}\,h^{2}}{27}}  
- { \frac {320\,d\,J^{2}\,g^{2}}{27}}    \nonumber \\
&&
 - { \frac {1568\,d^{2}\,J^{2}\,g}{243}}  
- { \frac {352\,g\,J^{2}\,h^{2}}{81}}  
+ { \frac {128\,J\,\Delta \,h^{2}\,d\,g}{9}} 
\end{eqnarray}

\begin{eqnarray} 
W_5 &=& 
{ \frac {298\,J^{2}\,d^{2}\,h^{2}}{243}}   
- { \frac {2728\,J^{2}\,\Delta ^{2}\,d^{2}\,h^{2}}{81}}  
+ { \frac {3656\,d\,J^{3}\,\Delta \,h^{2}}{243}}  
- { \frac {10256\,d\,J^{3}\,\Delta ^{3}\,h^{2}}{243}}  
- { \frac {1568\,J\,\Delta \,h^{2}\,d^{3}}{243}}   \nonumber \\
&&
 + { \frac {128\,J\,\Delta \,d\,h^{4}}{27}} 
 + { \frac {955\,d^{4}\,h^{2}}{972}}  
 + { \frac {91\,d^{2}\,h^{4}}{108}}  
 - { \frac {14827\,J^{6}\,\Delta ^{6}}{10935}}  
 + { \frac {3763\,J^{6}\,\Delta ^{4}}{1215}}  
 - { \frac {5959\,J^{2}\,d^{4}}{3645}}    \nonumber \\
&& 
 - { \frac {19\,J^{2}\,h^{4}}{81}}  
+ { \frac {4601\,d^{2}\,J^{4}}{3645}}  
+ { \frac {4861\,J^{6}\,\Delta ^{2}}{3645}}  
+ { \frac {49\,h^{2}\,J^{4}}{81}}  
- { \frac {1220\,d^{2}\,g^{4}}{243}} 
 - { \frac {100\,g^{4}\,h^{2}}{81}}     \nonumber \\
&& 
 + { \frac {2440\,J^{4}\,\Delta ^{2}\,h^{2}}{243}}  
+ { \frac {376\,d\,J^{5}\,\Delta ^{3}}{1215}}  
+ { \frac {11216\,J^{4}\,\Delta ^{2}\,d^{2}}{1215}}  
+ { \frac {92\,J^{2}\,\Delta ^{2}\,h^{4}}{27}} 
 + { \frac {1532\,J^{2}\,\Delta ^{2}\,d^{4}}{729}} \nonumber \\
&&
 - { \frac {4088\,d\,J^{5}\,\Delta }{3645}} 
 + { \frac {1397\,d^{6}}{8748}}  
 - { \frac {13\,h^{6}}{1620}}  + { \frac {4213\,J^{6}}{10935}} 
 + { \frac {808\,J^{3}\,\Delta \,d^{3}}{3645}} 
 - { \frac {1828\,J^{4}\,\Delta ^{4}\,d^{2}}{243}}  \nonumber \\
&& 
 - { \frac {3650\,J^{4}\,\Delta ^{4}\,h^{2}}{243}}  
+ { \frac {3776\,J^{2}\,\Delta ^{2}\,g^{2}\,d^{2}}{243}}  
+ { \frac {5152\,g\,J^{2}\,\Delta ^{2}\,d^{3}}{729}}  
- { \frac {544\,J^{2}\,\Delta ^{2}\,g^{2}\,h^{2}}{81}}   \nonumber \\
&&
 + { \frac {11008\,J^{2}\,\Delta ^{2}\,d\,g^{3}}{729}}  
- { \frac {2560\,J^{4}\,\Delta ^{4}\,g\,d}{243}}  
+ { \frac {1600\,J^{2}\,g\,d\,h^{2}}{243}} 
 + { \frac {26144\,d\,g\,J^{4}\,\Delta ^{2}}{729}} \nonumber \\
&&
 - { \frac {176\,J^{3}\,\Delta \,d^{2}\,g}{729}}  
+ { \frac {5968\,J^{3}\,\Delta \,h^{2}\,g}{243}}  
- { \frac {64\,J^{3}\,\Delta \,d\,g^{2}}{27}}  
+ { \frac {256\,J\,\Delta \,h^{2}\,g^{3}}{27}} \nonumber \\
&&
 - { \frac {7936\,J^{3}\,\Delta ^{3}\,h^{2}\,g}{243}}  
+ { \frac {320\,J\,\Delta \,g\,h^{4}}{81}}  
- { \frac {256\,g^{6}}{135}}  
- { \frac {3664\,J^{2}\,d^{2}\,g^{2}}{729}}  
+ { \frac {18640\,J^{4}\,\Delta ^{2}\,g^{2}}{729}}     \nonumber \\
&&
+ { \frac {704\,h^{2}\,g\,d^{3}}{243}}  
+ { \frac {16\,g^{3}\,d\,h^{2}}{27}} 
+ { \frac {128\,J^{2}\,\Delta ^{2}\,g^{4}}{243}}  
- { \frac {1184\,g\,J^{5}\,\Delta }{729}}  
- { \frac {592\,J^{5}\,\Delta ^{3}\,g}{729}}    \nonumber \\
&&
 + { \frac {512\,g\,d\,J^{4}}{729}}  
+ { \frac {496\,J^{2}\,g^{2}\,h^{2}}{81}}  
- { \frac {64\,J^{3}\,\Delta \,g^{3}}{27}}  
- { \frac {9472\,J^{2}\,d\,g^{3}}{729}}  
+ { \frac {56\,g^{2}\,J^{4}\,\Delta ^{4}}{243}}     \nonumber \\
&&
 - { \frac {448\,J^{2}\,d^{3}\,g}{81}}  
+ { \frac {104\,g\,d\,h^{4}}{81}}  
+ { \frac {214\,d^{2}\,h^{2}\,g^{2}}{81}}  
- { \frac {512\,d\,g^{5}}{135}}  + { \frac {256\,J^{2}\,g^{4}}{243}}  
+ { \frac {9656\,g^{2}\,J^{4}}{729}}    \nonumber \\
&&
 + { \frac {17\,g^{2}\,h^{4}}{81}}  
- { \frac {7696\,g^{3}\,d^{3}}{2187}} 
- { \frac {101\,g^{2}\,d^{4}}{243}}  
+ { \frac {2296\,g\,d^{5}}{3645}} 
 - { \frac {1376\,J^{2}\,\Delta ^{2}\,g\,d\,h^{2}}{27}}    \nonumber \\
&& 
+ { \frac {128\,d\,h^{2}\,J\,\Delta \,g^{2}}{27}}  
- { \frac {1088\,h^{2}\,d^{2}\,J\,\Delta \,g}{81}}
\end{eqnarray}

\begin{eqnarray} 
W_6 &=& 
{ \frac {20581\,J^{2}\,d^{5}}{10935}}  
- { \frac {1397\,h^{2}\,d^{5}}{2916}}     \nonumber \\
&& 
 + { \frac {371\,J^{7}\,\Delta ^{5}}{7290}} 
 + { \frac {1871\,J^{6}\,d}{1215}}  
 - { \frac {3064\,J\,\Delta \,h^{2}\,d^{4}}{729}}  
 + { \frac {63404\,J^{4}\,\Delta ^{4}\,h^{2}\,d}{729}} \nonumber \\
&& 
 + { \frac {25384\,J^{2}\,\Delta ^{2}\,h^{2}\,d^{3}}{729}}  
+ { \frac {72112\,J^{3}\,\Delta ^{3}\,h^{2}\,d^{2}}{729}}  
- { \frac {2492\,J^{2}\,\Delta ^{2}\,h^{4}\,d}{81}}  
- { \frac {1900\,J^{3}\,\Delta \,h^{2}\,d^{2}}{81}}     \nonumber \\
&& 
 - { \frac {36088\,J^{4}\,\Delta ^{2}\,h^{2}\,d}{729}}  
- 16\,J\,\Delta \,h^{4}\,d^{2} + { \frac {8\,J\,\Delta \,h^{6}}{45}}  
+ { \frac {J^{4}\,h^{2}\,d}{243}} 
 + { \frac {3874\,J^{3}\,\Delta \,d^{4}}{10935}}      \nonumber \\
&&  
 + { \frac {19426\,J^{5}\,\Delta \,d^{2}}{10935}}  
- { \frac {121261\,J^{6}\,\Delta ^{4}\,d}{10935}}  
+ { \frac {143\,J^{2}\,h^{4}\,d}{81}} 
 + { \frac {2222\,J^{2}\,h^{2}\,d^{3}}{729}} 
 - { \frac {841\,h^{4}\,d^{3}}{972}}     \nonumber \\
&&  
 + { \frac {11\,h^{6}\,d}{108}}  
+ { \frac {8341\,J^{7}\,\Delta }{10935}}  
+ { \frac {14456\,J^{7}\,\Delta ^{3}}{10935}} 
- { \frac {5689\,J^{4}\,d^{3}}{10935}}  
+ { \frac {84224\,J^{2}\,g^{5}}{3645}}  
+ { \frac {32989\,d^{5}\,g^{2}}{10935}}     \nonumber \\
&& 
+ { \frac {11588\,d^{4}\,g^{3}}{2187}}  
- { \frac {9424\,d^{2}\,g^{5}}{3645}}  
+ { \frac {256\,h^{2}\,g^{5}}{135}}  
+ { \frac {2212\,d^{3}\,g^{4}}{729}}  
+ { \frac {11176\,d^{6}\,g}{10935}}     \nonumber \\
&& 
 - { \frac {150584\,J^{6}\,g}{10935}}  
- { \frac {37408\,J^{4}\,g^{3}}{2187}}  
+ { \frac {20\,h^{4}\,g^{3}}{27}}  
+ { \frac {4\,h^{6}\,g}{45}}  - { \frac {2576\,d\,g^{6}}{405}}  
- { \frac {33004\,J^{6}\,\Delta ^{2}\,d}{10935}}     \nonumber \\
&& 
 + { \frac {4928\,J^{4}\,\Delta ^{2}\,d^{3}}{2187}}  
- { \frac {3416\,J^{4}\,\Delta ^{4}\,d^{3}}{729}}  
- { \frac {1208\,J^{3}\,\Delta ^{3}\,h^{4}}{81}}  
+ { \frac {1252\,J^{5}\,\Delta ^{3}\,d^{2}}{3645}}     \nonumber \\
&& 
 + { \frac {96854\,J^{5}\,\Delta ^{5}\,h^{2}}{3645}}  
- { \frac {266\,J^{5}\,\Delta \,h^{2}}{243}}  
+ { \frac {230\,J^{3}\,\Delta \,h^{4}}{81}}  
- { \frac {4636\,J^{2}\,\Delta ^{2}\,d^{5}}{729}} 
+ { \frac {8126\,J^{6}\,\Delta ^{6}\,d}{2187}}    \nonumber \\
&& 
 + { \frac {8201\,d^{7}}{43740}}  
- { \frac {18560\,J^{5}\,\Delta ^{3}\,h^{2}}{729}}  
- { \frac {256\,J\,\Delta \,h^{2}\,g^{4}}{243}}  
- { \frac {61568\,J^{2}\,\Delta ^{2}\,d^{3}\,g^{2}}{2187}}    \nonumber \\
&& 
 - { \frac {17192\,J^{2}\,\Delta ^{2}\,d^{4}\,g}{729}}  
+ { \frac {128\,J^{2}\,\Delta ^{2}\,d\,g^{4}}{9}}  
- { \frac {6656\,J^{2}\,\Delta ^{2}\,d^{2}\,g^{3}}{729}}  
+ { \frac {48640\,J^{4}\,\Delta ^{4}\,h^{2}\,g}{729}}    \nonumber \\
&& 
 - { \frac {19840\,J^{2}\,\Delta ^{2}\,h^{2}\,g^{3}}{729}}  
+ { \frac {5920\,J^{3}\,\Delta ^{3}\,h^{2}\,g^{2}}{243}}  
- { \frac {352\,J^{4}\,\Delta^{4}\,d^{2}\,g}{27}}  
- { \frac {16208\,J^{4}\,\Delta ^{4}\,d\,g^{2}}{729}}    \nonumber \\
&& 
 - { \frac {6136\,J^{2}\,\Delta ^{2}\,h^{4}\,g}{243}}  
+ { \frac {11080\,J^{5}\,\Delta \,d\,g}{2187}}  
- { \frac {37376\,J^{3}\,\Delta \,h^{2}\,g^{2}}{729}}  
+ { \frac {8432\,J^{3}\,\Delta \,d^{3}\,g}{10935}}     \nonumber \\
&& 
 + { \frac {4048\,J^{3}\,\Delta \,d^{2}\,g^{2}}{2187}}  
+ { \frac {256\,J^{3}\,\Delta \,d\,g^{3}}{729}} 
+ { \frac {6224\,J^{2}\,h^{2}\,d^{2}\,g}{729}}  
+ { \frac {5392\,J^{2}\,h^{2}\,d\,g^{2}}{729}}      \nonumber \\
&& 
 - { \frac {52576\,J^{4}\,\Delta ^{2}\,h^{2}\,g}{729}}  
- { \frac {56032\,J^{4}\,\Delta ^{2}\,d^{2}\,g}{3645}}  
- { \frac {68008\,J^{4}\,\Delta ^{2}\,d\,g^{2}}{2187}}  
- { \frac {1472\,J\,\Delta \,h^{4}\,g^{2}}{243}}   \nonumber \\
&& 
 + { \frac {944\,J^{5}\,\Delta ^{3}\,d\,g}{243}}  
+ { \frac {63280\,J^{2}\,d^{3}\,g^{2}}{2187}}  
+ { \frac {200\,J^{2}\,h^{4}\,g}{81}} 
 + { \frac {48968\,J^{2}\,d^{4}\,g}{3645}}  
 + { \frac {19576\,J^{5}\,\Delta \,g^{2}}{2187}}    \nonumber \\
&&  
 + { \frac {31744\,J^{2}\,d\,g^{4}}{729}}  
+ { \frac {1220\,h^{2}\,d\,g^{4}}{243}}  
+ { \frac {1396\,J^{6}\,\Delta ^{2}\,g}{405}}  
+ { \frac {23872\,J^{4}\,\Delta ^{2}\,g^{3}}{2187}} 
 + { \frac {8384\,J^{2}\,d^{2}\,g^{3}}{243}}    \nonumber \\
&&  
 + { \frac {6464\,J^{2}\,h^{2}\,g^{3}}{729}} 
 + { \frac {30176\,J^{4}\,d^{2}\,g}{10935}}  
 + { \frac {928\,J^{4}\,h^{2}\,g}{729}}  
 - { \frac {448\,h^{4}\,d^{2}\,g}{243}}  
 - { \frac {1148\,h^{2}\,d^{4}\,g}{729}}     \nonumber \\
&&  
 - { \frac {11\,h^{4}\,d\,g^{2}}{243}}  
- { \frac {22712\,J^{6}\,\Delta ^{4}\,g}{1215}} 
+ { \frac {26912\,J^{6}\,\Delta ^{6}\,g}{10935}} 
 + { \frac {64\,J^{3}\,\Delta \,g^{4}}{243}}  
 + { \frac {9788\,J^{5}\,\Delta ^{3}\,g^{2}}{2187}} \nonumber \\
&& 
 + { \frac {42112\,J^{2}\,\Delta ^{2}\,g^{5}}{3645}}  
+ { \frac {78424\,J^{4}\,d\,g^{2}}{2187}} 
+ { \frac {3848\,h^{2}\,d^{2}\,g^{3}}{729}}  
+ { \frac {202\,h^{2}\,d^{3}\,g^{2}}{243}}  
- { \frac {8224\,J^{4}\,\Delta ^{4}\,g^{3}}{729}}    \nonumber \\
&& 
 - { \frac {368\,g^{7}}{135}}  
+ { \frac {108544\,J^{3}\,\Delta ^{3}\,h^{2}\,d\,g}{729}}  
- { \frac {10304\,J\,\Delta \,h^{2}\,d^{3}\,g}{729}}  
- { \frac {7552\,J\,\Delta \,h^{2}\,d^{2}\,g^{2}}{243}}    \nonumber \\
&& 
 - { \frac {22016\,J\,\Delta \,h^{2}\,d\,g^{3}}{729}}  
+ { \frac {1216\,J^{2}\,\Delta ^{2}\,h^{2}\,d\,g^{2}}{243}}  
+ { \frac {18416\,J^{2}\,\Delta ^{2}\,h^{2}\,d^{2}\,g}{243}}    \nonumber \\
&& 
 - { \frac {6080\,J\,\Delta \,h^{4}\,d\,g}{243}}  
- { \frac {57488\,J^{3}\,\Delta \,h^{2}\,d\,g}{729}}
\end{eqnarray}
}
%

\section{High temperature expansion of the magnetization of the spin-$S$ 
XXZ model }  \label{B}

From the high temperature expansion of the Helmholtz free energy
$\mathit{W}_\mathit{S}$ of the irreducible spin-$S$ XXZ model in
Ref.\cite{spin-S}, we obtain the magnetization ${\cal M}$ as function
of the squared norm of  the  spin $s(s+1)$, up to
order $\beta^6$:

\begin{equation}\label{B1}
{\cal M} = - \frac{\partial W_S}{\partial h} 
	= M_1 \ \beta
	+ M_2 \ \beta^2
	+ M_3 \ \beta^3
	+ M_4 \ \beta^4
	+ M_5 \ \beta^5
	+ M_6 \ \beta^6
	+ {\cal O} (\beta^7)
\end{equation}
where the coefficients $M_j$ are polynomials in $s(s+1)$,

{\footnotesize
\begin{equation} 
M_1 = {\frac {h\,s\,(s + 1)}{3}}
\end{equation}

\begin{equation} 
M_2 = - \left({ \frac {4}{45}} \,h\,d 
	+ { \frac {2}{9}} \,\Delta \,h\right)\,s^{2}\,(s + 1)^{2} 
	+ { \frac {h\,d\,s\,(s + 1)}{15}}
\end{equation}

\begin{eqnarray} 
M_3 &=&  -  \left({ \frac {4}{135}} \,h - { \frac {8}{945}} \,h\,d^{2} 
 - { \frac {14}{135}} \,\Delta ^{2}\,h 
 - { \frac {16}{135}} \,\Delta \,h\,d\right)\,s^{3}\,(s + 1)^{3} 
                          \nonumber\\
&&
- \left({ \frac {1}{30}} \,h 
+ { \frac {4}{105}} \,h\,d^{2} 
+ { \frac {1}{45}} \,h^{3} 
+ { \frac {1}{45}} \,\Delta ^{2}\,h 
+ { \frac {4}{45}} \,\Delta \,h\,d\right)\,s^{2}\,(s + 1)^{2}  \nonumber\\
&&
 - \left( - { \frac {1}{42}} \,h\,d^{2} 
 + { \frac {1}{90}} \,h^{3}\right)\,s\,(s + 1)
\end{eqnarray}

\begin{eqnarray} 
M_4 &=& - \left( - { \frac {8}{4725}} \,h\,d 
+ { \frac {128}{4725}} \,\Delta \,h\,d^{2} 
+ { \frac {344}{4725}} \,\Delta ^{2}\,h\,d 
+ { \frac {92}{2025}} \,\Delta ^{3}\,h 
- { \frac {8}{225}} \,\Delta \,h
- { \frac {16}{14175}} \,h\,d^{3}\right)\,s^{4}\,(s + 1)^{4}    \nonumber \\
&&
 - \left( - { \frac {8}{135}} \,\Delta \,h^{3} 
- { \frac {352}{4725}} \,\Delta \,h\,d^{2}
- { \frac {16}{945}} \,h^{3}\,d 
- { \frac {32}{4725}} \,h\,d^{3} 
- { \frac {16}{675}} \,\Delta ^{3}\,h \right.   \nonumber \\
&& \left.
 - { \frac {386}{4725}} \,\Delta ^{2}\,h\,d
 - { \frac {16}{1575}} \,h\,d 
 - { \frac {19}{675}} \,\Delta \,h \right) s^{3}\,(s + 1)^{3}
  - \left({ \frac {1}{675}} \,\Delta ^{3}\,h 
 + { \frac {97}{4725}} \,h\,d^{3} 
 + { \frac {3}{350}} \,h\,d \right.        \nonumber\\
 && \left.
 + { \frac {2}{945}} \,h^{3}\,d 
 + { \frac {1}{75}} \,\Delta \,h 
 - { \frac {4}{135}} \,\Delta \,h^{3} 
 + { \frac {32}{1575}} \,\Delta ^{2}\,h\,d 
 + { \frac {64}{1575}} \,\Delta \,h\,d^{2} \right) s^{2}\,(s + 1)^{2}
                                 \nonumber\\
&&
 - \left({ \frac {1}{126}} \,h^{3}\,d 
 - { \frac {1}{90}} \,h\,d^{3}\right)\,s\,(s + 1)
\end{eqnarray}

\begin{eqnarray} 
M_5 &=&  -\left({ \frac {32}{2835}} \,\Delta \,h\,d 
 - { \frac {64}{1701}} \,\Delta ^{3}\,h\,d 
 - { \frac {718}{42525}} \,\Delta ^{4}\,h \right.  \nonumber\\
 &&\left.
 - { \frac {4}{2835}} \,h 
- { \frac {64}{42525}} \,\Delta \,h\,d^{3} 
- { \frac {64}{2835}} \,\Delta ^{2}\,h\,d^{2} 
+ { \frac {848}{42525}} \,\Delta ^{2}\,h 
+ { \frac {32}{93555}} \,h\,d^{4} 
- { \frac {32}{14175}} \,h\,d^{2}\right)         \nonumber\\
&&
s^{5}\,(s + 1)^{5}\mbox{} 
- \left( - { \frac {8}{14175}} \,h\,d^{2} 
- { \frac {16}{31185}} \,h\,d^{4} 
+ { \frac {8}{1575}} \,h^{3}\,d^{2} 
- { \frac {4}{567}} \,h 
- { \frac {4}{525}} \,h^{3}\right.     \nonumber\\
&&
 + { \frac {128}{14175}} \,\Delta \,h\,d 
 + { \frac {176}{2835}} \,\Delta ^{3}\,h\,d 
 + { \frac {352}{14175}} \,\Delta \,h\,d^{3} 
 + { \frac {64}{945}} \,\Delta ^{2}\,h\,d^{2} 
 + { \frac {32}{525}} \,\Delta \,h^{3}\,d   \nonumber\\
 && \left.
 + { \frac {383}{42525}} \,\Delta ^{2}\,h 
 + { \frac {229}{14175}} \,\Delta ^{4}\,h 
 + { \frac {358}{4725}} \,\Delta ^{2}\,h^{3}\right)s^{4}\,(s+ 1)^{4}
 - \left( - { \frac {11}{1575}} \,h\,d^{2}
 - { \frac {184}{31185}} \,h\,d^{4} \right.     \nonumber\\
 &&
 - { \frac {128}{14175}} \,h^{3}\,d^{2} 
 - { \frac {1}{210}} \,h 
 - { \frac {158}{14175}} \,h^{3} 
 - { \frac {2}{945}} \,h^{5} 
 - { \frac {38}{1575}} \,\Delta \,h\,d 
 - { \frac {16}{567}} \,\Delta ^{3}\,h\,d    \nonumber\\
 &&
 - { \frac {652}{14175}} \,\Delta \,h\,d^{3}
 - { \frac {8}{135}} \,\Delta ^{2}\,h\,d^{2} 
 - { \frac {136}{14175}} \,\Delta \,h^{3}\,d
 - { \frac {271}{14175}} \,\Delta ^{2}\,h 
 - { \frac {64}{14175}} \,\Delta ^{4}\,h    \nonumber\\
 && \left.
 + { \frac {424}{14175}} \,\Delta ^{2}\,h^{3}\right)s^{3}\,(s + 1)^{3}
  - \left({ \frac {41}{6300}} \,h\,d^{2} 
  + { \frac {1}{77}} \,h\,d^{4}
  - { \frac {11}{3150}} \,h^{3}\,d^{2} 
  + { \frac {1}{840}} \,h - { \frac {37}{6300}} \right. \,h^{3}
                                                  \nonumber\\
&&
 - { \frac {1}{630}} \,h^{5} 
 + { \frac {13}{1575}} \,\Delta \,h\,d 
 + { \frac {2}{945}} \,\Delta ^{3}\,h\,d 
 + { \frac {4}{189}} \,\Delta \,h\,d^{3} 
 + { \frac {1}{63}} \,\Delta ^{2}\,h\,d^{2} 
 - { \frac {128}{4725}} \,\Delta \,h^{3}\,d   \nonumber\\
 && \left.
+ { \frac {1}{315}} \,\Delta ^{2}\,h 
+ { \frac {1}{3780}} \,\Delta ^{4}\,h 
- { \frac {23}{4725}} \,\Delta ^{2}\,h^{3} \right) s^{2}\,(s+ 1)^{2}  
                          \nonumber\\
&&
 - \left( - { \frac {5}{792}} \,h\,d^{4} 
 - { \frac {1}{2520}} \,h^{5} 
 + { \frac {1}{180}} \,h^{3}\,d^{2}\right)\,s\,(s + 1)
\end{eqnarray}

\begin{eqnarray} 
M_6 &=&  - \left( - { \frac {1592}{1488375}} \,h\,d  \right.      \nonumber \\
 &&
 + { \frac {16064}{49116375}} \,h\,d^{3} 
 + { \frac {1472}{638512875}} \,h\,d^{5} 
 - { \frac {8336}{893025}} \,\Delta ^{2}\,h\,d 
 + { \frac {12032}{893025}} \,\Delta ^{3}\,h\,d^{2} 
                 \nonumber\\
&&
- { \frac {35072}{49116375}} \,\Delta \,h\,d^{4} 
+ { \frac {70424}{4465125}} \,\Delta ^{4}\,h\,d
+ { \frac {2528}{893025}} \,\Delta \,h 
- { \frac {1672}{178605}} \,\Delta ^{3}\,h 
+ { \frac {5132}{893025}} \,\Delta ^{5}\,h   \nonumber \\
&& \left.
+ { \frac {5792}{1964655}} \,\Delta ^{2}\,h\,d^{3} 
+ { \frac {1856}{1488375}} \,\Delta \,h\,d^{2} \right) s^{6}\,(s + 1)^{6}
- \left( - { \frac {8}{1488375}} \,h\,d 
- { \frac {304}{606375}} \,h\,d^{3}  \right.  \nonumber   \\
&&
+ { \frac {16}{4725}} \,h^{3}\,d 
+ { \frac {24512}{70945875}} \,h\,d^{5} 
- { \frac {32}{66825}} \,h^{3}\,d^{3} 
- { \frac {16}{893025}} \,\Delta ^{2}\,h\,d 
- { \frac {1696}{33075}} \,\Delta ^{3}\,h\,d^{2}   \nonumber \\
&&
- { \frac {44416}{16372125}} \,\Delta \,h\,d^{4} 
- { \frac {5942}{165375}} \,\Delta ^{4}\,h\,d
+ { \frac {7628}{893025}} \,\Delta \,h 
+ { \frac {2}{6615}} \,\Delta ^{3}\,h 
+ { \frac {64}{2835}} \,\Delta \,h^{3}    \nonumber  \\
&&
- { \frac {2656}{297675}} \,\Delta ^{5}\,h
- { \frac {992}{14175}} \,\Delta ^{3}\,h^{3} 
- { \frac {128}{4725}} \,\Delta \,h^{3}\,d^{2} 
- { \frac {1328}{14175}} \,\Delta ^{2}\,h^{3}\,d 
- { \frac {17152}{654885}} \,\Delta ^{2}\,h\,d^{3} 
             \nonumber   \\
&&\left.
+ { \frac {496}{165375}} \,\Delta \,h\,d^{2} \right) s^{5}\,(s + 1)^{5} 
- \left({ \frac {1259}{992250}} \,h\,d 
+ { \frac {1168}{606375}} \,h\,d^{3} 
+ { \frac {4}{675}} \,h^{3}\,d  \right.  \nonumber  \\
&&
+ { \frac {13136}{70945875}} \,h\,d^{5} 
+ { \frac {2096}{467775}} \,h^{3}\,d^{3} 
+ { \frac {4}{1575}} \,h^{5}\,d 
+ { \frac {1607}{59535}} \,\Delta ^{2}\,h\,d 
+ { \frac {2048}{33075}} \,\Delta ^{3}\,h\,d^{2}  \nonumber \\
&&
+ { \frac {112864}{5457375}} \,\Delta \,h\,d^{4} 
+ { \frac {13882}{496125}} \,\Delta ^{4}\,h\,d
+ { \frac {1543}{1190700}} \,\Delta \,h 
+ { \frac {17137}{1190700}} \,\Delta ^{3}\,h 
+ { \frac {139}{4725}} \,\Delta \,h^{3}    \nonumber  \\
&&
+ { \frac {457}{99225}} \,\Delta ^{5}\,h 
- { \frac {656}{42525}} \,\Delta ^{3}\,h^{3} 
+ { \frac {2}{175}} \,\Delta \,h^{5} 
+ { \frac {64}{1575}} \,\Delta \,h^{3}\,d^{2} 
+ { \frac {16}{525}} \,\Delta ^{2}\,h^{3}\,d    \nonumber  \\
&& \left.
+ { \frac {746}{13475}} \,\Delta ^{2}\,h\,d^{3} 
+ { \frac {2158}{165375}} \,\Delta \,h\,d^{2}\right)s^{4}\,(s + 1)^{4} 
- \left( - { \frac {253}{220500}} \,h\,d 
- { \frac {12071}{1819125}} \,h\,d^{3}  \right.  \nonumber  \\
&&
+ { \frac {1}{4725}} \,h^{3}\,d 
- { \frac {42166}{7882875}} \,h\,d^{5} 
- { \frac {194}{51975}} \,h^{3}\,d^{3} 
+ { \frac {2}{4725}} \,h^{5}\,d 
- { \frac {683}{33075}} \,\Delta ^{2}\,h\,d   \nonumber \\
&&
- { \frac {2606}{99225}} \,\Delta ^{3}\,h\,d^{2} 
- { \frac {165736}{5457375}} \,\Delta \,h\,d^{4} 
- { \frac {8111}{992250}} \,\Delta ^{4}\,h\,d 
- { \frac {193}{33075}} \,\Delta \,h 
- { \frac {481}{66150}} \,\Delta ^{3}\,h    \nonumber  \\
&&
+ { \frac {142}{14175}} \,\Delta \,h^{3} 
- { \frac {1}{1323}} \,\Delta ^{5}\,h 
+ { \frac {152}{14175}} \,\Delta ^{3}\,h^{3} 
+ { \frac {44}{4725}} \,\Delta \,h^{5} 
+ { \frac {184}{14175}} \,\Delta \,h^{3}\,d^{2}   \nonumber \\
&& \left.
+ { \frac {563}{14175}} \,\Delta ^{2}\,h^{3}\,d 
- { \frac {48809}{1091475}} \,\Delta ^{2}\,h\,d^{3} 
- { \frac {2423}{110250}} \,\Delta \,h\,d^{2}\right)s^{3}\,(s + 1)^{3}
- \left({ \frac {43}{88200}} \,h\,d \right.    \nonumber  \\
&&
+ { \frac {647}{161700}} \,h\,d^{3} 
- { \frac {31}{6300}} \,h^{3}\,d 
+ { \frac {177571}{18918900}} \,h\,d^{5} 
- { \frac {32}{6237}} \,h^{3}\,d^{3} 
- { \frac {19}{18900}} \,h^{5}\,d     \nonumber  \\
&&
+ { \frac {4}{1225}} \,\Delta ^{2}\,h\,d 
+ { \frac {74}{33075}} \,\Delta ^{3}\,h\,d^{2} 
+ { \frac {4553}{363825}} \,\Delta \,h\,d^{4} 
+ { \frac {37}{66150}} \,\Delta ^{4}\,h\,d 
+ { \frac {13}{14700}} \,\Delta \,h    \nonumber \\
&&
+ { \frac {17}{22050}} \,\Delta ^{3}\,h 
- { \frac {11}{3150}} \,\Delta \,h^{3} 
+ { \frac {1}{26460}} \,\Delta ^{5}\,h 
- { \frac {2}{2835}} \,\Delta ^{3}\,h^{3} 
+ { \frac {11}{4725}} \,\Delta \,h^{5} 
- { \frac {4}{189}} \,\Delta \,h^{3}\,d^{2}   \nonumber  \\
&& \left.
- { \frac {1}{135}} \,\Delta ^{2}\,h^{3}\,d
+ { \frac {4553}{363825}} \,\Delta ^{2}\,h\,d^{3}
+ { \frac {11}{1470}} \,\Delta \,h\,d^{2}\right)s^{2}\,(s+1)^{2}    \nonumber  \\
&&
- \left({ \frac {5}{1188}} \,h^{3}\,d^{3} 
- { \frac {1}{1800}} \,h^{5}\,d 
- { \frac {691}{163800}} \,h\,d^{5}\right)\,s\,(s + 1)
\end{eqnarray}
}


\end{document}